\newcommand{\be}{\begin{equation}}
\newcommand{\ee}{\end{equation}}
\newcommand {\gapprox}
   {\raisebox{-0.7ex}{$\stackrel {\textstyle>}{\sim}$}}
\newcommand {\lapprox}
   {\raisebox{-0.7ex}{$\stackrel {\textstyle<}{\sim}$}}
\newcommand{\beeq}{\begin{eqnarray}}
\newcommand{\eeeq}{\end{eqnarray}}
\newcommand{\diii}{\frac{{\rm d}^3\sigma_{ep\rightarrow eXR_c}}{{\rm d}\beta {\rm d}Q^2{\rm d}x_{I\!\!P}}}
\newcommand{\diiii}{\frac{{\rm d}^4\sigma_{ep\rightarrow eXR_c}}{{\rm d}\beta {\rm d}Q^2{\rm d}x_{I\!\!P}{\rm d}t}}

\documentstyle[12pt,epsfig]{article}
\newlength{\dinwidth}
\newlength{\dinmargin}
\begin{document}
\titlepage
hep-ph/9602285 \hfill INP 1715/PH
\begin{center}
{\large \bf QCD: Quantum Chromodynamic Diffraction}\footnote{
From talks given in the diffractive session at the
Workshop on Proton, Photon and Pomeron Structure, Durham, September 1995}
\vskip 0.5cm
{\large K.~Golec--Biernat$^2$, J.~P.~Phillips$^3$}
\vskip 0.5cm
{\it  
$^2$ \footnotesize{H. Niewodnicza\'nski Institute of Nuclear Physics,
Radzikowskiego 152, 31-342 Krak\'ow, Poland}\\
$^3$ \footnotesize{Department of Physics, University of Liverpool,
Liverpool. L69 3BX.}\\}
\vskip1cm
\end{center}
\begin{abstract}
  The first measurements of the diffractive structure function $F_2^{D(3)}$
  at HERA are discussed. A factorisable interpretation in which a 
  partonic structure is assigned to the pomeron is investigated through 
  QCD analyses in which both the quark and gluon densities are 
  permitted to vary freely. A method of measuring the longitudinal 
  structure function of the pomeron without changing the $ep$ centre of 
  mass energy is presented. The possibility that the pomeron structure may
  receive a large contribution from gluons, relative to quarks, at high 
  $\beta$ is highlighted, and the experimental signatures which may 
  support such a structure are reviewed.
\end{abstract}
\vskip -3cm
\section{Introduction}
 The phenomenology of high energy diffraction succeeds in 
 correlating many of the features of high energy hadron--hadron 
 interactions in terms of relatively few parameters. However since 
 the advent of QCD as the theory of hadronic physics, it has been 
 clear that ultimately there must be an understanding of this 
 phenomenology in terms of QCD. With the advent of the $ep$ collider 
 HERA, for the first time it has become possible to probe the 
 regions of proton structure involved in the diffractive interactions which 
 form the bulk of the proton interaction cross section at high energy.  
 
 Recently the H1 and ZEUS collaborations 
 have quantified the diffractive contribution to the proton structure function
 by measuring the ``diffractive structure function'' $F_2^{D(3)}$ 
 (see \cite{H11,ZEUS2} for details of the analyses).  
 These measurements are made as a function of three kinematic 
 variables, $x$, $\beta$ and $x_{I\!\!P}$, or equivalently 
 $\beta$, $Q^2$ and $x_{I\!\!P}$, which are defined as follows:
\begin{equation}
x=\frac {-q^2}{2P\cdot q}\,\,\,\,\,\,
\,\,\,\,\,\,\,\,\,
x_{I\!\!P} = \frac{q\cdot (P-P')}{q\cdot P}\,\,\,\,\,\,
\,\,\,\,\,\,\,\,\,
Q^2 = -q^2\,\,\,\,\,\,
\,\,\,\,\,\,\,\,\,\,\,
\beta = \frac{-q^2}{2q\cdot (P-P')} .
          \label{definition}
\end{equation}
  Here $q$, $P$ and $P'$ are the $4$--momenta of the virtual boson, 
  the incident proton, and the final state colourless remnant $R_c$ 
  (proton or low mass excited state) respectively~\cite{H11,ZEUS2}. 
  Note that $x=\beta x_{I\!\!P}$. A structure function is  
  defined~\cite{H11} in analogy with the decomposition of the 
  unpolarised total $ep$ cross section:
\begin{equation}
\frac{{\rm d}^4 \sigma_{ep\rightarrow eXR_c}}{{\rm d}x{\rm d}
  Q^2{\rm d}x_{I\!\!P}dt}
  = \frac{4\pi \alpha_{em}^2}{xQ^4}\,
  \left\{1-y+\frac{y^2}{2[1+R^{D(4)}(x,Q^2,x_{I\!\!P},t)]}\right\}
  \,F_2^{D(4)}(x,Q^2,x_{I\!\!P},t)
           \label{eq:defF2D}
\end{equation}
  where $X$ is the hadronic system excluding the colourless remnant $R_c$. 
  For these measurements, no accurate determination of $t=(P-P^{\prime})^2$ 
  was possible, and the measured structure function $F_2^{D(3)}$  was 
  evaluated from the differential cross section 
  $\diii=\int \diiii\,{\rm d}t$ such that
\begin{equation}
\diii = \frac{4\pi\alpha_{em}^2}{\beta Q^4}\left\{ 1-y + \frac{y^2}{2}
\right\} F_2^{D(3)}(\beta,Q^2,x_{I\!\!P})
\end{equation}
  where $R^{D(4)}$ is set to 0 for all $t$ following the original 
  procedure of~\cite{Ingprytz1}. An excellent fit to all data points, 
  irrespective of $\beta$ and $Q^2$,
  is obtained assuming a dependence $x_{I\!\!P}^{-n}$ with a single 
  exponent (H1: $n=1.19\pm0.06(stat.)\pm0.07(syst.)$, $94\%$C.L., 
  ZEUS: $n=1.30\pm0.08(stat.)^{+0.08}_{-0.14}(syst.)$). Such a universal
  dependence is expected na\"{\i}vely if the diffractive deep--inelastic
  process involves the interaction of a virtual photon with a (colourless)
  target in the incident proton whose characteristics are not dependent on
  $x_{I\!\!P}$, and which carries only a small fraction of the proton's 
  momentum. Furthermore, the values for $n$ are consistent with 
  that expected if the diffractive mechanism may be encapsulated in the 
  parameterisation that describes  ``soft hadronic" diffractive
  interactions, namely the pomeron ($I\!\!P$) with 
  $\alpha(t)=\alpha_{I\!\!P}(0)+\alpha' t$ and $\alpha_{I\!\!P}(0)=1.085$, 
  $\alpha'=0.25\,{\rm GeV}^{-2}$~\cite{ppdiff,Softpom}.
  The diffractive structure function may therefore be written in the
  ``factorisable'' form
\begin{equation}
 F_2^{D(3)}(\beta,Q^2,x_{I\!\!P}) = f(x_{I\!\!P}) F_2^{I\!\!P}(\beta,Q^2)
 \label{eq:xpfactor}
\end{equation}
  with $f(x_{I\!\!P}) \propto x_{I\!\!P}^{-n}$, 
  and the natural interpretation of $F_2^{I\!\!P}(\beta,Q^2)$ is then 
  that of the deep-inelastic structure of the diffractive exchange 
  ($I\!\!P$).

\section{QCD Analysis of the H1 Data}
\begin{figure}[tb] \unitlength 1mm
\centering
\begin{picture}(150,80)
\put(0,0){\epsfig{figure=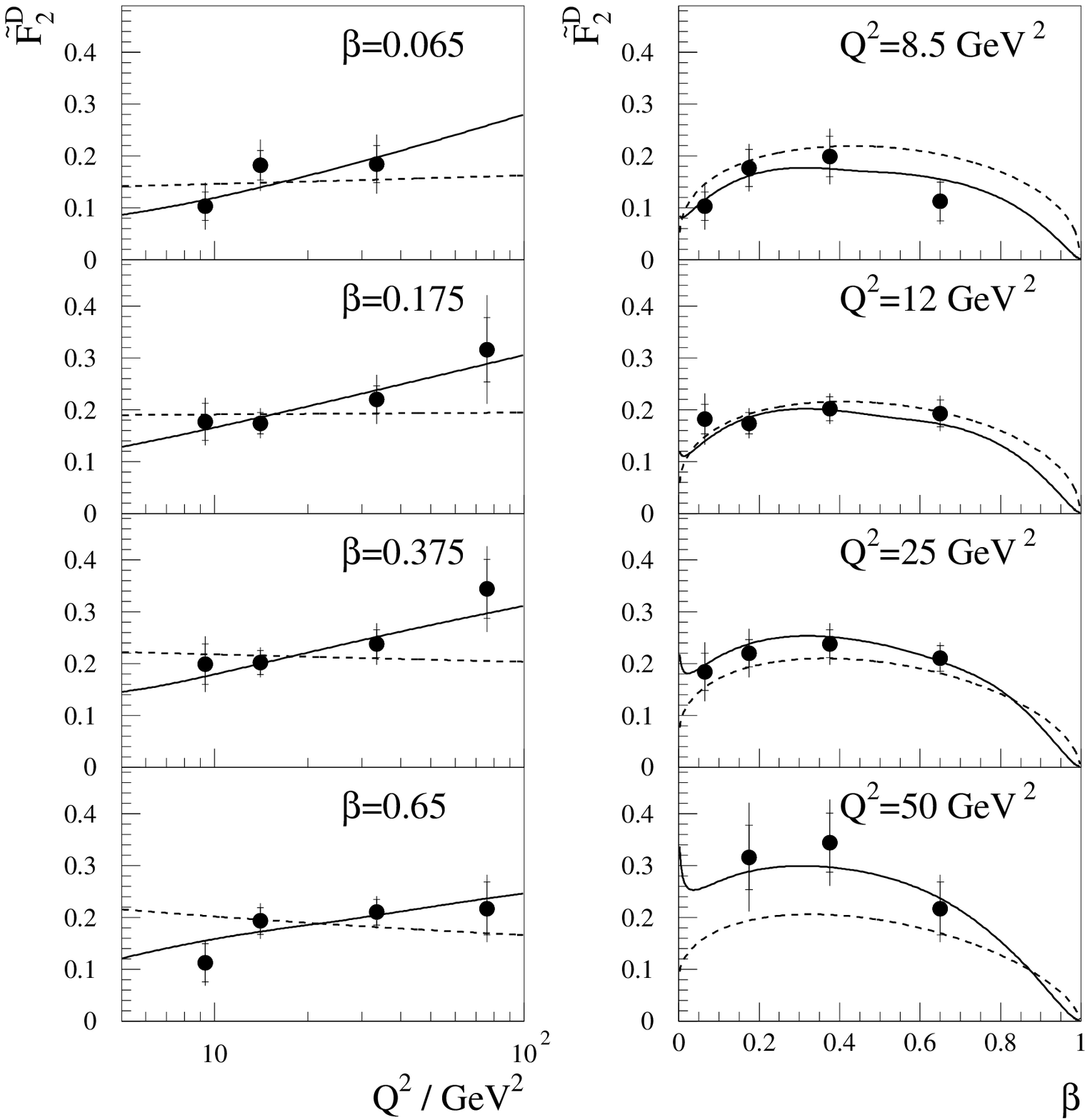,width=0.5\textwidth}}
\put(52,-2){\epsfig{figure=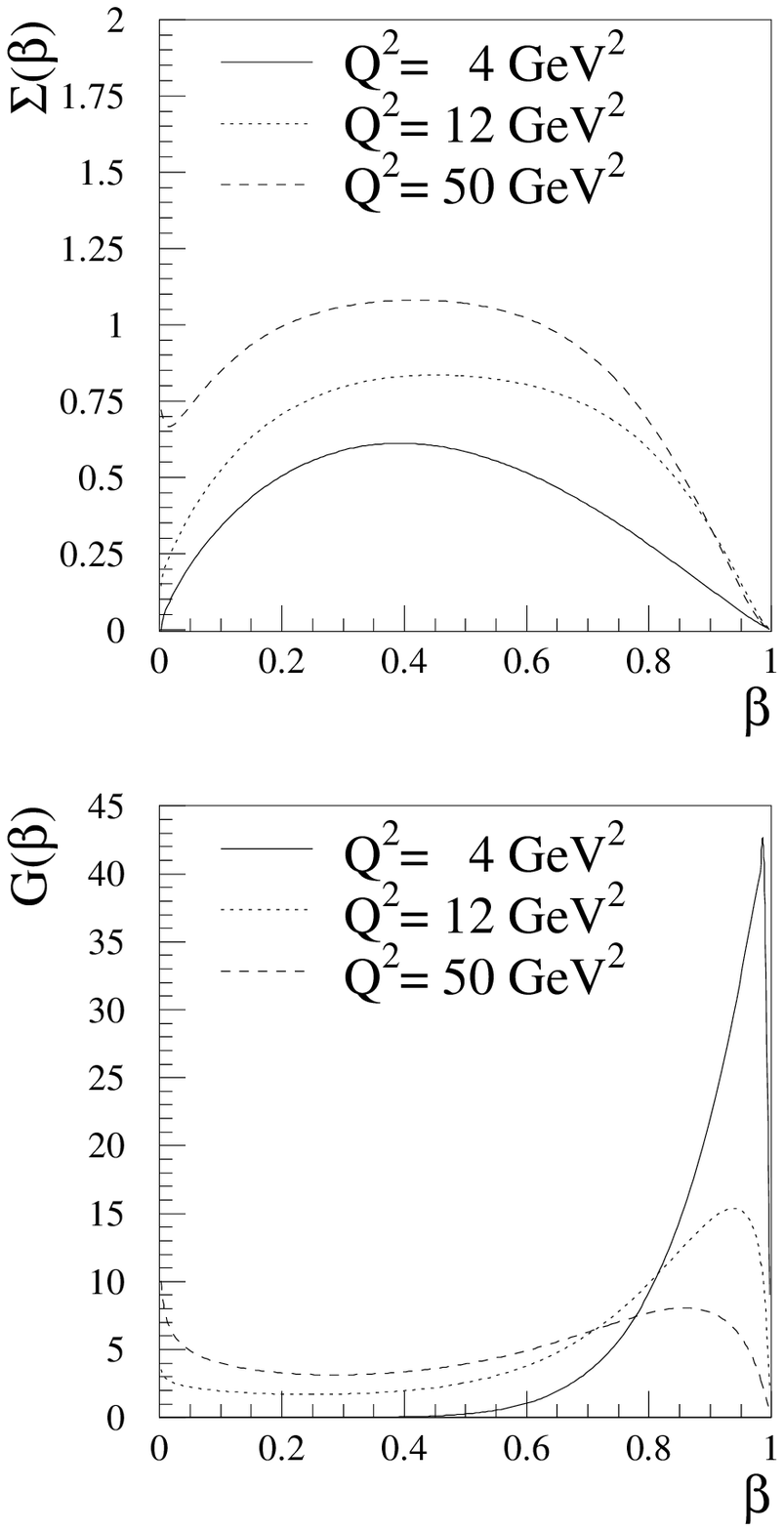,width=0.5166\textwidth}}
\put(5,0){a)}
\put(50,0){b)}
\put(100,45){c)}
\put(100,0){d)}
\end{picture}
    \caption{Dependence of $\tilde{F_2^D}$
     on $Q^2$ (a) and $\beta$ (b); superimposed are the results of two LO 
     $\log Q^2$ DGLAP QCD fits. The dashed line shows a fit in 
     which at the starting scale, $Q^2=4\,{\rm GeV^2}$, diffraction
     is attributed to the exchange of only quarks ($\chi^2/{\rm dof}$ 
     of $13/12$, $37\%{\rm C.L.}$). The solid line shows 
     a fit in which both quarks and gluons may contribute to the 
     diffractive mechanism at $Q^2=4\,{\rm GeV^2}$
     ($\chi^2/{\rm dof}$ = $4/9$, $91\%{\rm C.L.}$). The data are those
     published in~$\rm [2]$. The fitted quark singlet (c) and gluon (d) 
     densities as a function of $\beta$ for different values of $Q^2$.}
    \label{fig:H1Fit}
\vspace*{-0.3cm}
\end{figure}

  In order to investigate the $\beta$ and $Q^2$ dependencies of 
  $F_2^{D(3)}$ it is convenient to define an integral of 
  $F_2^{D(3)}$ over a fixed range in $x_{I\!\!P}$ 
  which, assuming the simple factorisation of (\ref{eq:xpfactor}), is 
  proportional to the structure function of the $I\!\!P$:
\begin{equation}
  \tilde{F}_2^D(\beta,Q^2) = \int^{x_{I\!\!P_H}}_{x_{I\!\!P_L}}
  F_2^{D(3)}(\beta,Q^2,x_{I\!\!P})\,{\rm d}x_{I\!\!P} =
  F_2^{I\!\!P}(\beta,Q^2) \cdot \int_{x_{I\!\!P_L}}^{x_{I\!\!P_H}}
  f(x_{I\!\!P})\,{\rm d}x_{I\!\!P}
\end{equation}
  This procedure 
  avoids the need to specify the theoretically ill-defined normalisation
  of the diffractive flux, and permits direct comparison of the data 
  with any theoretical model. 
  The H1 data for $\tilde{F}_2^D(\beta,Q^2)$ 
  are shown in figure \ref{fig:H1Fit}.
  The lack of any substantial dependence of $\tilde{F}_2^D(\beta,Q^2)$ 
  on $Q^2$ and the substantial contribution at $\beta<1$ 
  suggests that the structure resolved by the electron in high $Q^2$ 
  diffractive interactions may be partonic in origin~\cite{H11}. 
  The interpretation of $\beta$ is then that of the appropriate 
  Bjorken scaling variable in deep--inelastic diffraction.
  This hypothesis may be tested by introducing parton 
  densities for the pomeron such that (in the leading ${\rm log}Q^2$ 
  approximation)
\begin{equation}
  F_2^{I\!\!P}(\beta,Q^2) = \sum_{i=1,n}e_i^2 \beta \left[
  q_i(\beta,Q^2) + \bar{q}_{i}(\beta,Q^2) \right]
  \label{eq:pomparton}
\end{equation}
  where $q_{i}(\beta,Q^2)$ are the density functions for the $n$ quark
  flavours considered. The evolution of these parton density functions 
  with $Q^2$ may then be calculated using the DGLAP evolution equations,
  and the range of possible solutions for the parton densities 
  which are compatible with the data investigated by a standard QCD 
  fit procedure.
  
  However, it has been observed that 
  the lack of any fall of $\tilde{F}_2^D$ with 
  increasing $Q^2$ at large $\beta$ contrasts with the violations of 
  scale invariance exhibited by the structure function of a typical 
  hadron~\cite{H11}. For example, the proton structure function $F_2$ 
  rises with increasing $Q^2$ for $x\,\lapprox\,0.15$, and falls with 
  increasing $Q^2$ for $x\,\gapprox\,0.15$. This decrease at
  high $x$ is an inevitable consequence of DGLAP QCD evolution for an 
  object with a structure built from the evolution of ``valence'' quarks 
  in which the latter predominate at large Bjorken--$x$. By analogy, a 
  structure function for which any violation of scale invariance amounts 
  to an increase with $\log Q^2$ at high Bjorken--$x$ ($\beta$), is likely 
  to have a large gluon, relative to quark, parton density at high $\beta$.
  
  To quantify any substantive evidence for a gluonic contribution to 
  $\tilde{F}_2^D$, H1 performed an analysis in which the DGLAP evolution 
  equations were solved numerically in the leading (LO) and next-to-leading
  order (NLO) ${\rm log}Q^2$ approximations~\cite{FITPARIS}. 
  Starting from a scale $Q^2_0=4\,{\rm GeV^2}$, the following 
  forms were assumed for the quark flavour 
  singlet\footnote{A non--singlet contribution is not 
  considered as a $0^+$ exchange is assumed, despite the fact that meson
  exchange contributions to the measured cross section are possible.
  This assumption is supported insofar as the data 
  for $F_2^{D(3)}$ may be parameterisd by a single trajectory with 
  intercept close to unity.} distribution
  $\sum(\beta)=\sum_i \beta\left[q_i(\beta) + \bar{q_i}(\beta) \right]$
  and gluon distribution $G(\beta)=\beta g(\beta)$ at an initial scale of
  $Q^2_0=4\,{\rm GeV^2}$:
\begin{equation}
\sum(\beta) = A_1 \beta^{A_2}(1-\beta)^{A_3} 
\end{equation}
\begin{equation}
G(\beta) = B_1 \beta^{B_2}(1-\beta)^{B_3} 
\end{equation}
  No momentum sum rule was imposed.
  These parton densities were evolved to higher $Q^2$ 
  and compared with the measurements of the $\beta$ dependence of 
  $\tilde{F}_2^D$ taking into account fully both the statistical and 
  systematic contributions to the uncertainty in the measured data.
  The results of two LO fits are shown, superimposed 
  on the data, in figure \ref{fig:H1Fit}. Though a solution in which 
  only quarks contribute to the pomeron structure at the starting scale
  (dashed line) does not {\em qualitatively} reproduce the rise of 
  $\tilde{F}_2^D$ with $Q^2$ at all values of $\beta$, this solution 
  provides a statistically acceptable description of the data
  ($\chi^2/{\rm dof}=13/12$, 37\%C.L.)   The addition of a 
  gluon density at $Q^2_0$ results in an excellent description 
  of $\tilde{F}_2^D(\beta,Q^2)$ ($\chi^2/{\rm dof}$ = $4/9$, 
  $91\%{\rm C.L.}$) which reproduces the rise with 
  $\log Q^2$ at higher $\beta$. In this solution, shown by the solid curve in
  figure~\ref{fig:H1Fit}, at $Q^2_0=4\,{\rm GeV^2}$ the gluons carry 
  $\sim 90\%$ of the momentum of diffractive exchange, and the fitted 
  gluon density is very hard, tending to $\beta=1$, indicating 
  that the structure of diffraction may involve the leading exchange 
  of a single gluon~(figure~\ref{fig:H1Fit}c,d). Repeating the analysis 
  at NLO reduces somewhat the fraction of the momentum carried by gluons.

\section{$F_L^{I\!\!P}$ and Violations of Factorisation}
  A NLO analysis allows a prediction for the longitudinal structure 
  function of the pomeron, $F_L^{I\!\!P}$, to be calculated. This prediction
  may then be tested directly against the data since the wide range in 
  $x_{I\!\!P}$ accessible at HERA results in a large variation in $eI\!\!P$
  centre of mass energy. Thus for a genuinely factorisable cross section 
  of the form $(\ref{eq:xpfactor})$ with 
  $R^{I\!\!P}=F_L^{I\!\!P}/(F_2^{I\!\!P}-F_L^{I\!\!P})$ greater than 0, 
  the $F_2^{D(3)}$ extracted from the data with the assumption that 
  $R^{I\!\!P}=0$ will be modified from the expectation of the factorisable
  expression () by a multiplicative factor 
  $\psi (\beta,Q^2,y)$ where
\begin{equation}
  \psi (\beta,Q^2,y) = \left[ 2(1-y)+\frac{y^2}{1+R^{I\!\!P}(\beta,Q^2)} 
\right] \,
 / \, \left[2(1-y)+y^2\right]
\end{equation}
  Thus $F_L^{I\!\!P}$ may be extracted directly from the data at fixed
  $ep$ centre of mass energy (in contrast to $F_L^p$) by measuring such 
  apparent deviations from factorisation. The correspondence 
  between $F_L^{I\!\!P}$ extracted in this way, and the prediction 
  of a QCD analysis would allow the validity of this factorisable 
  approach to be tested at NLO. 

\section{Global QCD Analysis of H1 and ZEUS Data}
\begin{figure}[htb]
   \vspace*{-1cm}
    \centerline{
     \psfig{figure=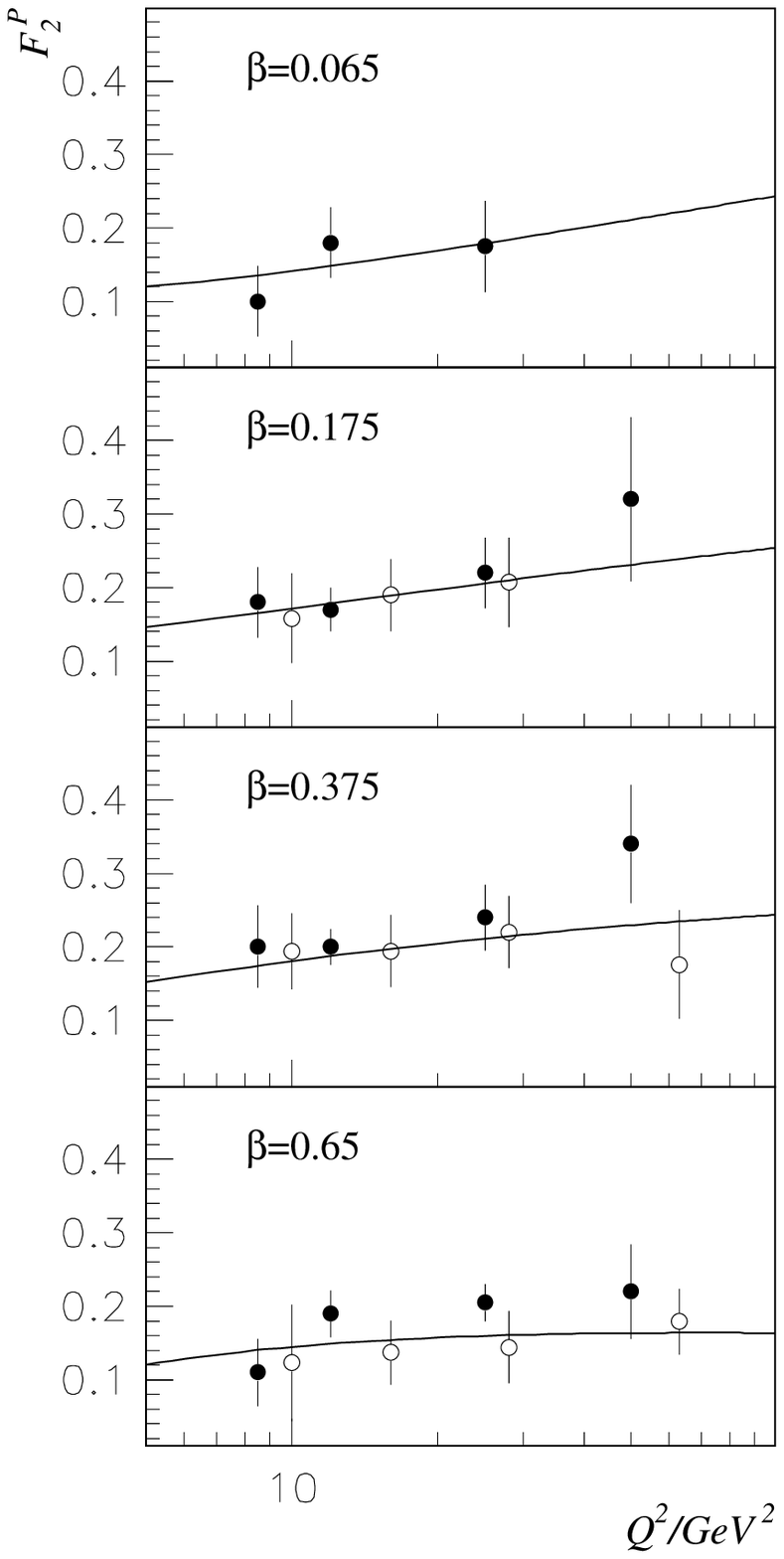,height=10cm,width=7cm}
     \psfig{figure=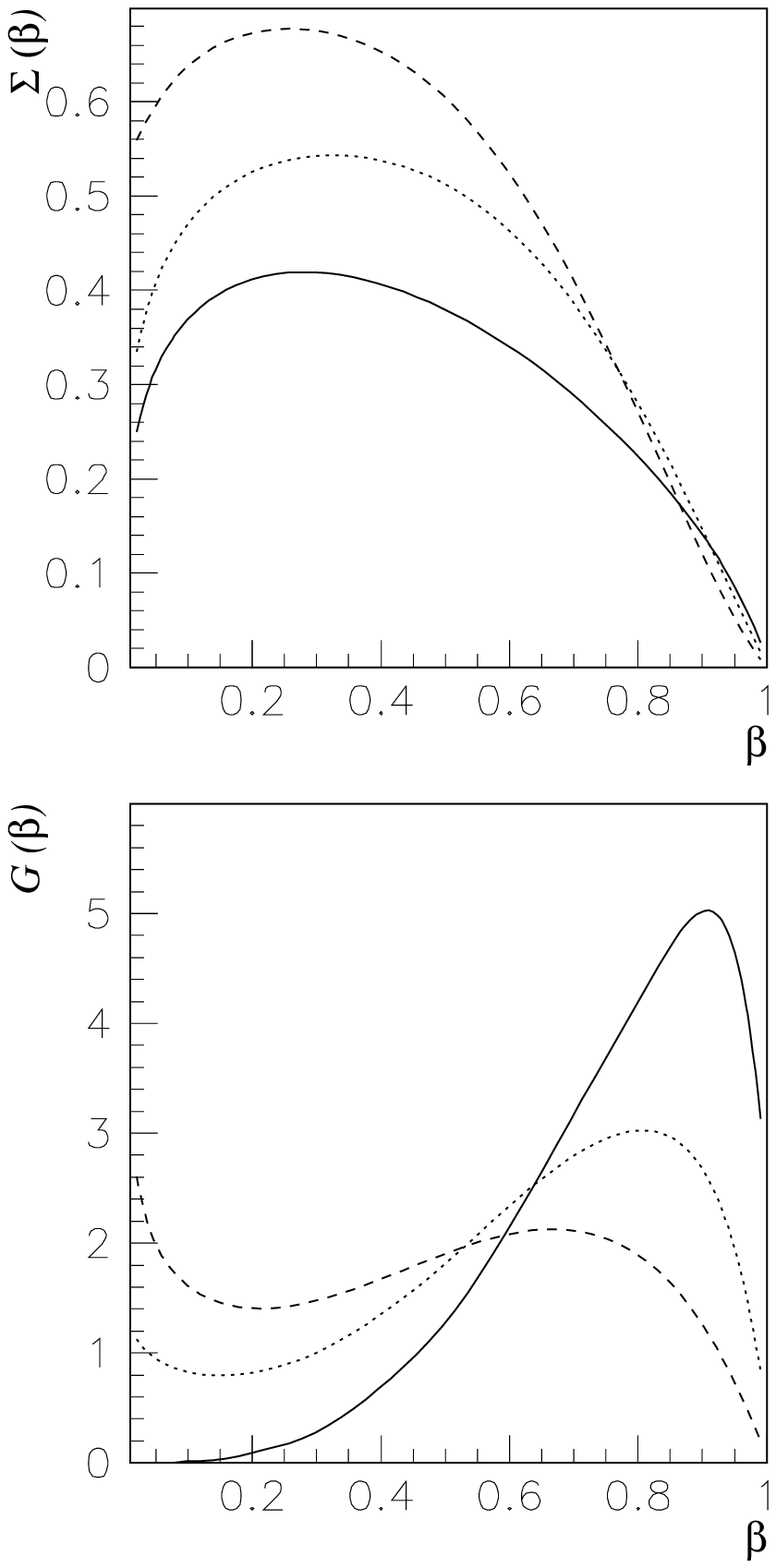,height=10cm,width=7cm}
     }
    \vspace*{-0.5cm}
     \caption{  
({\it Left}) - $F_2^P(\beta,Q^2)$ from H1 ({\it closed circles}) and
ZEUS ({\it open circles}) measurements compared to  the Fit 1 result
({\it solid line}). ZEUS data were rescaled by factor 1.76. 
({\it Right}) -
parton distributions in the pomeron from Fit 1 of Table 1 
for  $Q^2=4$ ({\it solid}),
12 ({\it dotted}) and 50 $GeV^2$({\it dashed line}). 
   \label{fig:golec}}
\end{figure}

  At this point we extend the H1 analysis by considering both the 
  H1~\cite{H11} and ZEUS~\cite{ZEUS2} data. A 
  leading log combined fit to these two 
  measurements of $F_2^{D(3)}(\beta,Q^2,x_{I\!\!P})$ is performed 
  assuming the factorisable form (\ref{eq:xpfactor})
  with with $f(x_{I\!\!P})=x_{I\!\!P}^{-n}$ and permitting the 
  parameterisation of the parton densities at the starting scale $Q_0^2$ 
  and the exponent $n$ to vary simultaneously. Table \ref{tab:fitpar}
  summarises the results of this study, which are compared with the data
  in figure~\ref{fig:golec}. In the first fit
  all seven parameters are allowed to vary whilst $\Lambda$ is fixed at
  $200\,{\rm MeV}$. The second fit illustrates how the results change 
  when $\Lambda$ is set to $255\,{\rm MeV}$. Fits 3 and 4 show how the 
  parameters change if the value of $n$ is fixed to an intermediate value,
  and to the value determined by ZEUS respectively. The obtained 
  $\chi^2/{\rm dof}$ confirms observation of H1 and ZEUS that the 
  present data are in a very good agreement with the factorisable 
  form (\ref{eq:xpfactor}) of $F_2^{D(3)}$. The fitted parameter $n$  
  is in an excellent agreement with  the H1 value ($n=-1.19$), 
  although  fixing it to the ZEUS value ($n=-1.30$) in the fit does 
  not lead to a significant deterioration of the fit quality.

\setcounter{table}{0}
\begin{table} 
\begin{center}
\vspace*{-1.cm}
{\small
\begin{tabular}{|c|r|r|r|r|}
\hline
Parameters & Fit 1 & Fit 2 & Fit 3 & Fit 4 \\
\hline
\hline
$-n$ &~~1.1791~~&~~~1.1794~~&~~1.25 {\it (fix)}   &~~1.30 {\it (fix)}  \\
\hline
$A_1$ & 0.066~~& 0.068~~& 0.034~~& 0.021~~\\
\hline
$A_2$ &  0.29~~& 0.31~~& 0.19~~&  0.11~~\\
\hline
$A_3$ & 0.72~~& 0.76~~& 0.79~~&  0.84~~\\
\hline
$B_1$ & 1.22~~& 1.01~~&  1.05~~&  0.72~~\\
\hline
$B_2$ & 3.13~~& 3.22~~& 2.90~~&  2.51~~\\
\hline
$B_3$ &  0.31~~& 0.21~~& 0.34~~&  0.27~~\\
\hline
\hline
$\Lambda $ &  ~~~200~~~~& ~~~255~~~~
                & ~~~200~~~~&  ~~~200~~~~   \\
\hline
$\chi^2/dof $ & 114/96~~& 114/96~~& 116/96~~& 120/96~~\\
\hline
\end{tabular}}
\caption{\it Fit results to $F_2^D(3)$ diffractive structure function data
           from H1 and ZEUS experiments. $\Lambda_{QCD}$ is fixed in all
             fits. The parameterisation is identical to that used in the 
             H1 fit, and is described in equations (7) and (8). Only
            statistical errors were taken into account in the fits.
         \label{tab:fitpar}}
\vskip -1cm
\end{center}
\end{table}

  The form of the fitted quark flavour singlet and gluon density functions
  are shown as a function of $\beta$ for different values of $Q^2$ in 
  figure \ref{fig:golec}. 
  The normalisation of the pomeron flux is somewhat arbitrary: any constant
  factor may be shifted between the pomeron flux and structure function in
  (\ref{eq:xpfactor}).  Here the pomeron flux parameterisation of 
  Berger et. al.~\cite{BERGER} and Streng~\cite{STRENG} was used to define
  absolutely the normalisation of parton densities in the pomeron. With this 
  normalisation the momentum sum 
  $\int_0^1{\rm d}\beta \Sigma(\beta)+G(\beta)$ is $1.7$ and the contribution
  to this sum from gluons is $85\%$ at $Q^2=4\,{\rm GeV^2}$ falling to 
  $75\%$ at $Q^2=200\,{\rm GeV^2}$, in agreement with the H1 analysis.
  The form of the gluon density at the starting 
  scale $Q^2_0=4\,{\rm GeV}^2$ is similar to that obtained in the H1 
  analysis, confirming the conclusion that the persistence of a rise of
  $F_2^{D(3)}$  with ${\rm log}Q^2$ can only be reproduced by a large 
  gluon (relative to quark) distribution at large $\beta$. However, it is
  worth repeating that the demand for such a gluon density cannot be 
  demonstrated conclusively given the magnitude of the total errors of
  the measurements. A plethora of gluon distributions have 
  been shown to yield a reasonable description of the HERA 
  data~\cite{GS,GK,FITPARIS,CK2} and so in the absence of more 
  accurate data the gluon density in the pomeron remains essentially 
  unconstrained. 

\section{Summary and Outlook}
  The analyses discussed here demonstrate that the H1 and ZEUS data 
  are compatible with a factorisable interpretation~\cite{IngSchl}
  in which the deep--inelastic diffractive structure is governed by a 
  partonic structure for the pomeron. 
  Whether such an approach provides more than 
  merely a compact parameterisation of the $F_2^{D(3)}$ data relies 
  crucially on the universality of these parton densities. Theoretically, 
  the collinear factorisation supporting such universality is predicted 
  to break down in diffractive jet production in hadron--hadron 
  collisions, but the mechanism responsible for this breakdown is not 
  expected in $ep$ scattering~\cite{CFS}. Although, factorisation has 
  not been proven rigorously for any diffractive 
  process~\cite{BERSOP,DELDUCA}, at this workshop
  there was general agreement that there is some justification for the 
  leading ${\rm log}Q^2$ approximation for $\beta < 1$. 

  Experimentally, the goals are clear. More accurate measurements of 
  $F_2^{D(3)}$ over as wide a kinematic range as possible are essential. 
  The QCD analyses suggest that the region
  of high $\beta$ is of great theoretical interest, where the potentially
  large gluon density may give rise to a large $F_L^{I\!\!P}$.
  Measurements at very high $y$ are sensitive to the 
  longitudinal component of the cross section and will test the validity 
  of DGLAP evolution beyond the leading ${\rm log}Q^2$ approximation.
  Measurements at $x_{I\!\!P}>0.05$ will establish whether 
  additional exchanges $(f,\pi,\rho,...)$ contribute to the production of 
  large rapidity gaps at HERA.   The universality of the parton densities 
  extracted from QCD analyses such as those presented here can be 
  tested directly by exclusive measurements. The wide range in gluon 
  densities compatible with the HERA data provide a wide range of predictions
  for the production of inclusive charm and high $E_T$ jets, and the 
  possibility that the presence of an additional hard scale 
  ($m_c$, high $E_T$, high $t$) could change the  energy ($x_{I\!\!P}$) 
  dependence  should be investigated. In conclusion, the simple 
  factorisable model in which 
  universal parton densities are ascribed to the pomeron has fulfilled the 
  most basic requirement of describing the inclusive diffractive cross 
  section, but it has yet to be tested in any substantial way. 

\section*{Acknowledgments}
It is a pleasure to thank the organizers for creating such a 
stimulating atmosphere during the workshop. One of us (K.G.) is grateful
for many fruitful discussions with H.Abramowicz and 
J.Kwieci\'nski. J.P. wishes to thank the other convenors of 
the diffractive session for support during an interesting week.
This research has been supported in part by KBN grant 
No.2P03B 231 08 and Maria Sklodowska-Curie Fund II (No.PAA/NSF-94-158). 


\end{document}